\documentclass[a4paper,11pt]{article}
\usepackage{jinstpub} 
\usepackage{siunitx}
\usepackage{lineno}

\newcommand{\dhyphen}{\mbox{--}}

\title{Performance of the first FBK DC-RSD1 production}

\author[a]{A.~Cassese$^{*}$}
\author[b,c]{R.~Arcidiacono}
\author[d]{G.~Bardelli} 
\author[e,f]{M.~Boscardin} 
\author[c]{N.~Cartiglia} 
\author[e,f]{M.~Centis~Vignali} 
\author[g]{T.~Croci} 
\author[h]{M.~Durando} 
\author[c]{S.~Durando} 
\author[c]{M.~Ferrero}
\author[i,d]{A.~Fondacci} 
\author[e,f]{O.~Hammad~Ali} 
\author[h]{L.~Lanteri} 
\author[d]{A.~Losana} 
\author[c,d]{L.~Menzio} 
\author[g]{A.~Morozzi} 
\author[j,g]{F.~Moscatelli} 
\author[i,g]{D.~Passeri} 
\author[e,f]{G.~Paternoster} 
\author[a]{G.~Sguazzoni} 
\author[c]{F.~Siviero} 
\author[h,c]{V.~Sola} 
\author[a]{L.~Viliani}

\affiliation[a]{Istituto Nazionale di Fisica Nucleare, Sezione di Firenze, Via Bruno Rossi 1, 50019 Sesto Fiorentino (FI), Italy}
\affiliation[b]{Universit\`a degli Studi del Piemonte Orientale, largo G. Donegani 2/3,28100 Novara, Italy}
\affiliation[c]{Istituto Nazionale di Fisica Nucleare, Sezione di Torino, via P. Giuria 1, 10125 Torino, Italy}
\affiliation[d]{CERN, Esplanade des Particules 1, 1211 Geneva 23, Switzerland} \affiliation[e]{Fondazione Bruno Kessler, via Sommarive 18, 38123 Povo (TN), Italy}
\affiliation[f]{TIFPA-INFN, via Sommarive 18, 38123 Povo (TN), Italy}
\affiliation[g]{Istituto Nazionale di Fisica Nucleare, Sezione di Perugia, via A. Pascoli 23c, 06123 Perugia, Italy}
\affiliation[h]{Universit\`a degli Studi di Torino, via P. Giuria 1, 10125 Torino, Italy}
\affiliation[i]{Universit\`a degli Studi di Perugia, via G. Duranti 93, 06125 Perugia, Italy}
\affiliation[j]{CNR-IOM, Sede secondaria di Perugia, via A. Pascoli 23c, 06123 Perugia, Italy}

\emailAdd{antonio.cassese@fi.infn.it}
\note[*]{Corresponding author.}

\abstract{
The DC Resistive Silicon Detector (DC-RSD) is based on thin LGAD sensors with a resistive DC-coupled readout, providing internal gain and intrinsic charge sharing.
This architecture enables excellent 4D tracking capabilities while maintaining reduced granularity.

The first production of such innovative devices, DC-RSD1, comprises matrices of elementary cells with varying geometries and sizes, as well as different technological implementations of the electrode layout and resistive layer.
This diversity allows for a systematic investigation of device performance, particularly in terms of charge sharing and effective gain as a function of fabrication parameters and design specifications.

This contribution presents a summary of the results obtained from the FBK DC-RSD1 production, including spatial and temporal resolution, and response uniformity across different sensor designs.
Measurements were carried out in three test beam campaigns at DESY and complemented by dedicated laboratory studies.
The devices were also subjected to irradiation campaigns to begin assessing their radiation hardness.
}

\keywords{Particle tracking detectors (Solid-state detectors); Solid state detectors; Timing detectors;}

\begin{document}
\maketitle
\flushbottom

\section{Introduction}
\label{sec:intro}
Future particle tracking detectors require excellent spatial and timing resolution while keeping the number of readout channels, power consumption,  and material budget under control.
In conventional segmented silicon detectors, the achievable spatial resolution is strongly correlated with the electrode pitch, motivating the development of increasingly fine detector segmentation.
This approach, however, comes at the cost of an increased channel density, readout complexity and power consumption.

Resistive Silicon Detectors (RSDs) combine the intrinsic gain of Low Gain Avalanche Diodes (LGADs) with controlled charge sharing in a resistive layer, enabling the simultaneous extraction of precise position and timing information from signals collected on neighbuoring electrodes.
Spatial resolutions of a few percent of the electrode pitch can be achieved while maintaining a channel density significantly lower than that of conventional pixel detectors.
Furthermore, the availability of timing information from multiple electrodes enables excellent time resolution, making RSDs attractive candidates for future 4D tracking applications~\cite{CARTIGLIA2022167228}.

Despite their excellent spatial and timing performance, the first RSD sensors, designed with AC-coupled electrodes, exhibit some non-ideal features~\cite{Fondacci_2025}.
The signal induced by a particle hit can spread over a large and position-dependent number of electrodes, leading to non-uniform charge sharing and increased reconstruction complexity.
Furthermore, the AC-coupled architecture produces bipolar signals with long discharge tails and relies on edge-only leakage current collection, which may become critical in large-area devices and after irradiation.
To address these limitations, a DC-coupled Resistive Silicon Detector (DC-RSD) architecture has been developed~\cite{MOSCATELLI2024169380,MENZIO2022167374}.
In this design, signal sharing is confined within isolated detector cells through the use of trench structures, while dedicated low-resistance paths enable direct signal collection at the readout electrodes.
The resulting unipolar signals exhibit shorter duration and improved confinement, providing enhanced performance uniformity and scalability.

The first FBK DC-RSD production (DC-RSD1) was completed in 2024 and includes several process variations aimed at optimising charge sharing, gain and signal collection.
This paper presents the design of the DC-RSD1 production together with the results of laboratory and beam-test measurements, with particular emphasis on the achieved spatial and timing resolution.

\section{DC-RSD concept and first production}
\label{sec:dcrsd}
\begin{figure}[ht]
\centering
  \includegraphics[width = 1.0\textwidth]{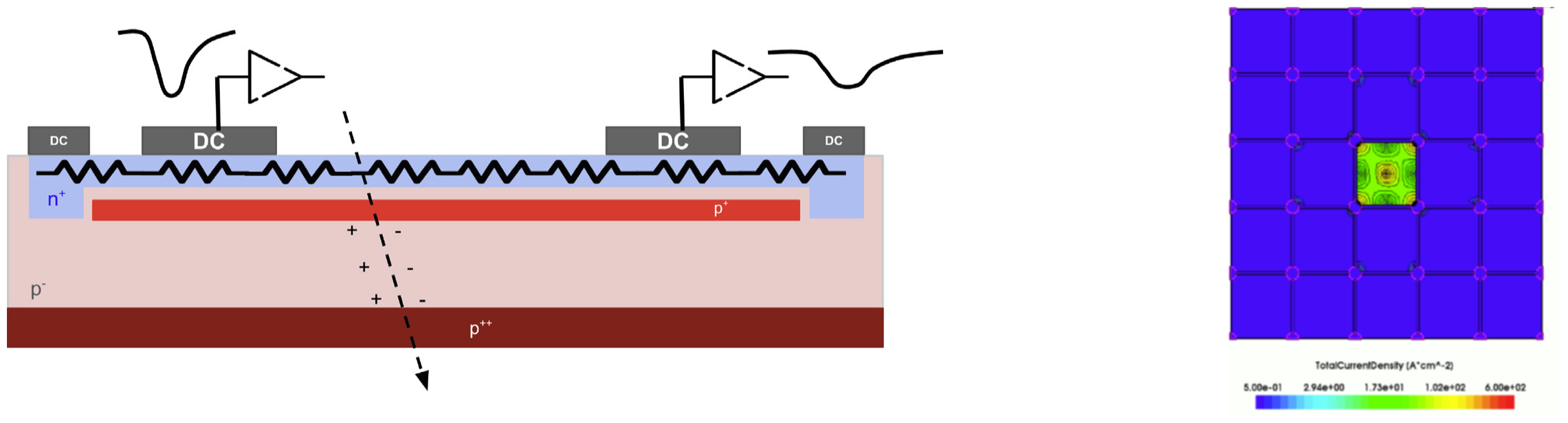}\\
  \caption{Schematic cross-section of the DC-RSD concept (left) and simulated surface current density for a particle crossing the centre of a pixel (right). The trench isolation confines signal propagation within the pixel area while preserving charge sharing among the readout electrodes.}
  \label{fig:layout}
\end{figure}

The DC-RSD architecture was developed to retain the excellent spatial and timing performance of the RSDs while addressing some of their main limitations.
The detector is based on a thin LGAD structure that features a continuous gain layer and a moderately resistive n$^+$ layer.
The charge sharing is confined within isolated detector cells defined by trench structures surrounding each pixel, as illustrated in figure~\ref{fig:layout}.
The trench isolation provides a controlled charge-sharing region, preventing signal propagation over a large number of neighbouring electrodes.
At the same time, the low-resistance paths offered by the resistive layer enable the electrode direct collection of DC signals.
As a result, the detector produces fast unipolar signals with improved charge confinement, local leakage current collection, and enhanced scalability towards large-area implementations.
For a particle crossing the centre of a pixel, TCAD simulations predict a strong confinement of the signal within the detector cell, as shown in figure~\ref{fig:layout}.

\begin{figure}[ht]
\centering
  \includegraphics[width = 1.0\textwidth]{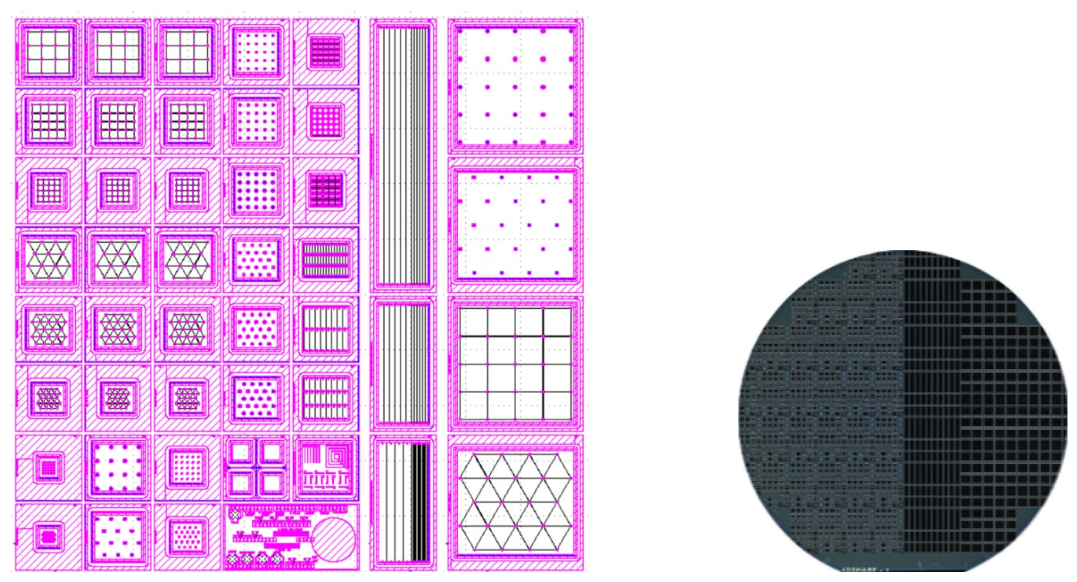}\\
  \caption{Reticle of the FBK DC-RSD1 production. The wafer includes several detector layouts and technological process splits designed to investigate the impact of the resistive layer, gain-layer dose and contact implementation on detector performance.}
  \label{fig:reticle}
\end{figure}

The first DC-RSD production (DC-RSD1) was completed at FBK in 2024~\cite{Centis_Vignali_2025} as a proof-of-concept implementation of the DC-RSD architecture.
Charge confinement was achieved through trench isolation structures, defining the geometry and dimensions of the detector cells.
The layout of the production is shown in figure~\ref{fig:reticle}.

The production includes several process splits designed to investigate the impact of key technological parameters on the detector performance.
These variations comprise three resistive n$^+$ layer implementations, two gain-layer doses and two electrode-to-silicon contact schemes.
Several pixel layouts with different geometries and pitches were included in the production in order to explore the design parameter space and identify the most promising configurations.

This work focuses on a selected subset of devices, corresponding to the most promising designs identified through preliminary laboratory characterisation.
Unirradiated sensors were studied using both laser-based measurements and beam tests performed at DESY and CERN.
The selected structures include square matrices with pitches ranging from 300~\si{\micro\meter} to 1000~\si{\micro\meter}, which constitute the main subject of the performance results presented in this paper.

\section{Experimental setup and data analysis}
\label{sec:setup}
\begin{figure}
    \centering
    \includegraphics[width = 0.5\textwidth]{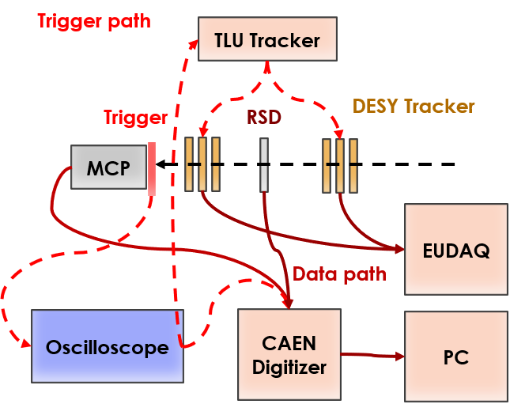}\\
    \caption{Beam-test setup showing the DUT, the tracking telescope used for spatial reconstruction, and the MCP detector used as timing reference. The figure highlights the trigger distribution and data acquisition paths used during the measurements.}
    \label{fig:setup}
\end{figure}

The performance of DC-RSD1 sensors was investigated through a combination of laboratory measurements and beam tests~\cite{ARCIDIACONO2025170796}.
A schematic representation of the experimental setup used for the beam-test measurements is shown in figure~\ref{fig:setup}.
The measurement setup included a tracking telescope~\cite{Jansen2016,Liu_2023} for spatial reconstruction and an MCP detector used as timing reference.
One of the fabricated DC-RSD1 samples was used as the device under test (DUT).
The results presented in this paper are primarily based on test-beam campaigns performed at the DESY-II facility using electron beams and at the CERN SPS using proton beams.
Signals from the detector electrodes were amplified using dedicated front-end electronics and digitised at multi-GHz sampling rates.
The hit position was reconstructed using an external tracking telescope, while the timing performance was evaluated with respect to an independent time reference detector.

\subsection{Position reconstruction}
\begin{figure}
    \centering
    \includegraphics[width = 1.0\textwidth]{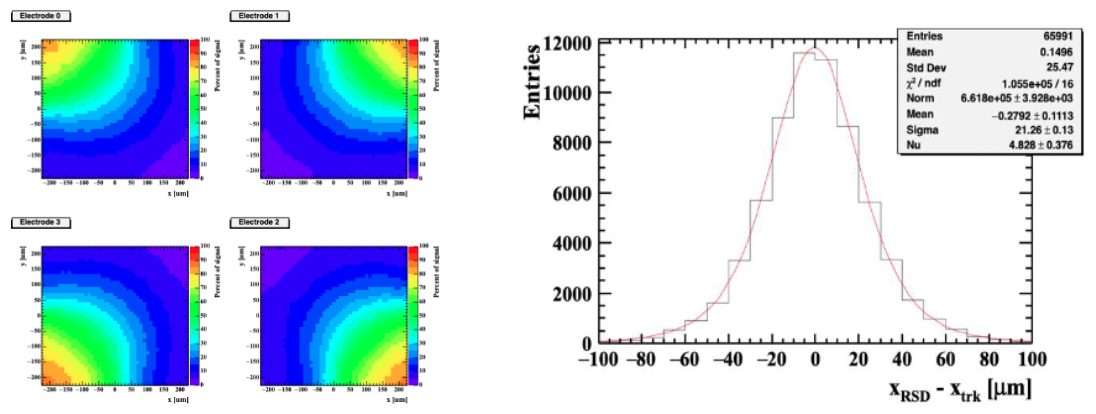}\\
        \caption{Left: signal fraction measured on each electrode as a function of the particle impact position for a square DC-RSD matrix. These maps are used to construct the lookup tables employed by the template-based position reconstruction algorithm. Right: example of the residual distribution between the reconstructed hit position and the external tracker prediction. The position resolution is extracted from a Student's-\emph{t} fit to the residual distribution.}
    \label{fig:spatial}
\end{figure}

The hit position was reconstructed exploiting the charge sharing among the electrodes surrounding the particle impact point.
For square pixel matrices, the reconstruction was based on the signals collected by the four electrodes defining the pixel cell.
A template-based approach was adopted~\cite{MENZIO2024169526,ARCIDIACONO2023,TORNAGO2021}.
Lookup tables describing the expected signal fractions as a function of the hit position were constructed from beam-test data acquired at the optimal operating bias.
An example of the signal-sharing templates for a 500~\si{\micro\meter} pitch square matrix is shown in figure~\ref{fig:spatial}.
For each event, the measured charge-sharing pattern was compared with the templates and the position corresponding to the best agreement was assigned to the hit.
The spatial resolution was evaluated from the residual distribution between the reconstructed position and the tracker prediction.
An example of the fitted residual distribution is also shown in figure~\ref{fig:spatial}.
The distributions were fitted using Student's-\emph{t} functions in order to account for non-Gaussian tails.

\subsection{Timing reconstruction}
\label{sec:timing}
\begin{figure}
    \centering
    \includegraphics[width = 1.0\textwidth]{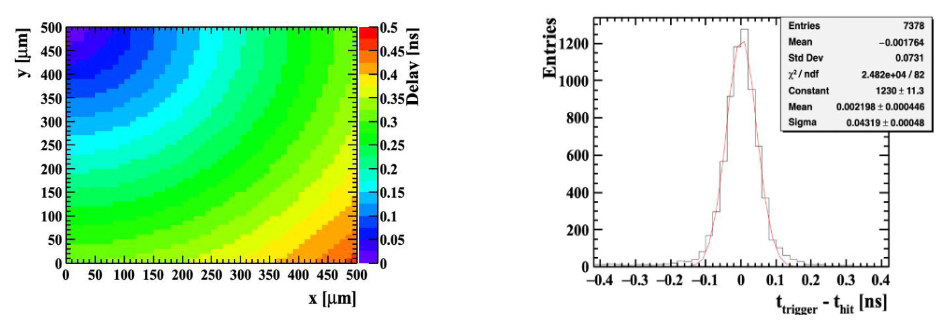}\\
        \caption{Left: signal propagation delay measured across a 500~\si{\micro\meter} pitch DC-RSD pixel matrix. The delay map is used to correct for position-dependent propagation effects in the timing reconstruction. Right: example of the residual distribution between the reconstructed particle time and the reference detector measurement. The timing resolution is extracted from a Gaussian fit to the distribution.}
    \label{fig:timing}
\end{figure}

The particle time of arrival was reconstructed by combining the timing information provided by the electrodes participating in the signal sharing.
The signal propagation time in the resistive layer depends on the hit position within the detector cell and was evaluated directly from the data.
An example of the measured propagation delay map for a 500~\si{\micro\meter} pitch matrix is shown in figure~\ref{fig:timing}.
The measured times of arrival were corrected accordingly, together with channel-dependent timing offsets.
The final particle timestamp was obtained from the weighted average of the corrected measurements, using the squared signal amplitudes as weights~\cite{Cartiglia2024}.
The timing resolution was extracted from the distribution of the difference between the reconstructed particle time and the reference detector measurement after subtracting the reference contribution in quadrature.

\section{Spatial resolution measurements}
\label{sec:spatial_resolution}
\begin{figure}
    \centering
    \includegraphics[width = 0.7\textwidth]{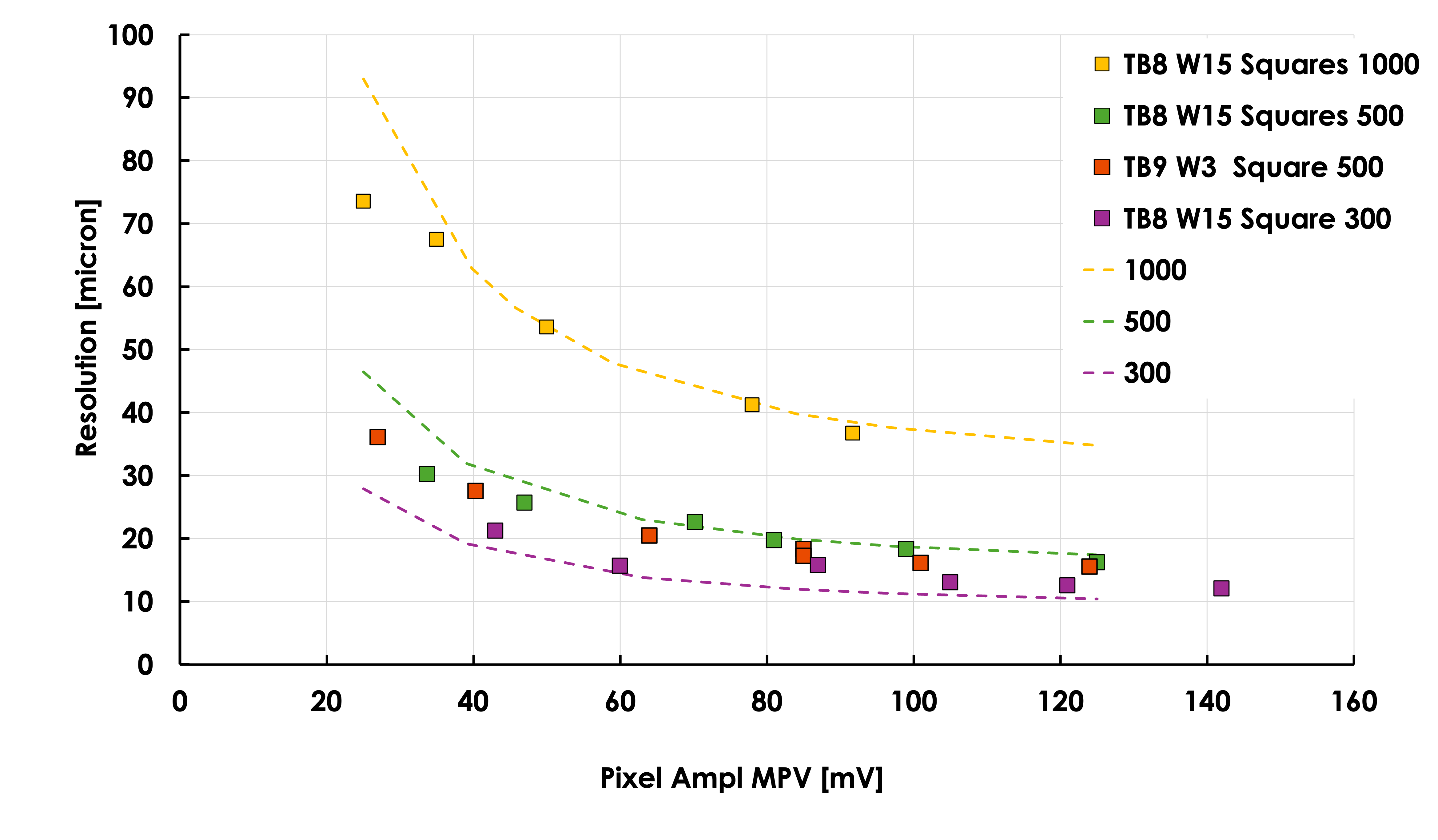}\\
    \caption{Position resolution as a function of the pixel signal amplitude for square DC-RSD matrices with pitches of 300, 500 and 1000~\si{\micro\meter}. The tracking telescope contribution has been subtracted in quadrature. The fit is a simple analytical model based on ideal resistive charge sharing and electronic noise. The model reproduces the observed behaviour using a single effective noise parameter.}
    \label{fig:model}
\end{figure}

The spatial resolution of the DC-RSD1 sensors was evaluated using the template-based reconstruction method described in the previous section.
Measurements were performed on square pixel matrices with pitches of 300, 500 and 1000~\si{\micro\meter} operated at different bias voltages, corresponding to different gain values.
For each detector configuration, the spatial resolution was extracted from the residual distribution after subtracting the contribution of the tracking telescope in quadrature.
The measured position resolution is shown in figure~\ref{fig:model} as a function of the most probable signal amplitude measured by the pixel (sum of the amplitudes seen by the four electrodes surrounding a pixel).

All investigated sensor geometries exhibit a similar behaviour.
The position resolution improves with increasing signal amplitude, reflecting the improved signal-to-noise ratio.
At sufficiently high gain, all sensors achieve a spatial resolution in the range of 3\dhyphen4\% of the electrode pitch.
This behaviour is observed over a wide range of pixel dimensions, indicating that the charge sharing mechanism scales effectively with the detector geometry.
The measured position resolution of DC-RSD sensors can be parametrized as the quadratic sum of a noise-like term, which decreases with increasing signal amplitude, and a constant contribution~\cite{ARCIDIACONO2023}:
\begin{equation}
 \sigma_x = pitch\times\sqrt{\left( \dfrac{k_1}{MPV}\right)^2+ k_2^2}
\end{equation}
where MPV is the most-probable signal amplitude, while the fitted parameters are the noise term $k_1 = 2.2$ \si{\milli\volt} and the constant term $k_2 = 0.03$.
As shown in figure~\ref{fig:model}, a single noise parameter provides a satisfactory description of all measured configurations.

\begin{figure}
    \centering
    \includegraphics[width = 0.7\textwidth]{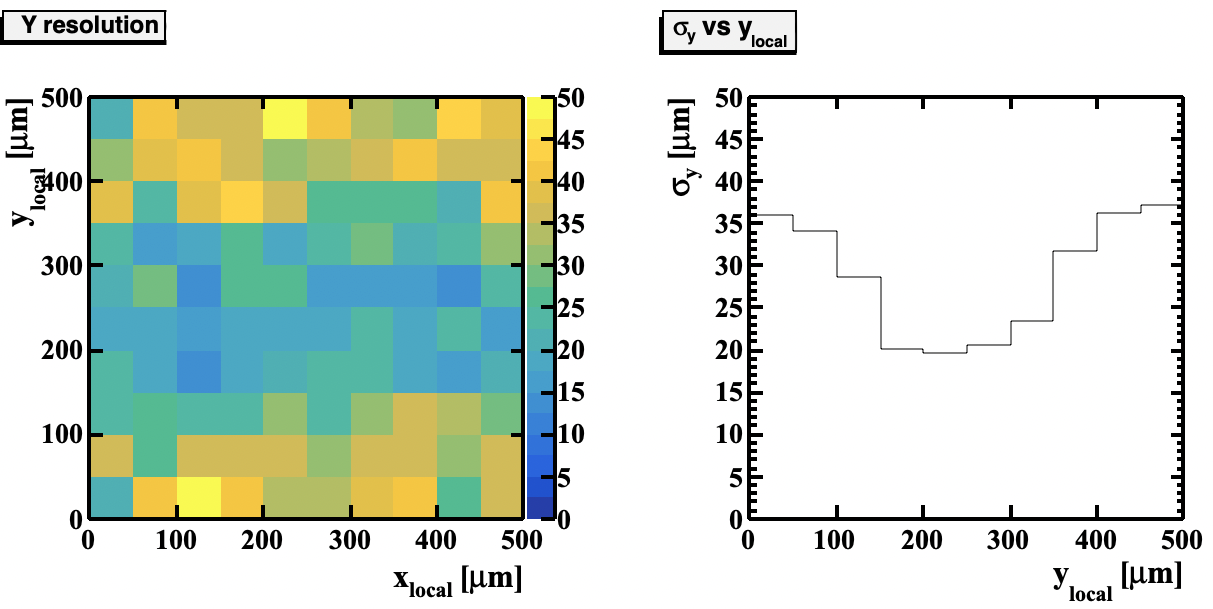}\\
    \caption{Spatial resolution uniformity across the pixel area. Left: two-dimensional map of the spatial resolution across the pixel area (all pixels superimposed). Right: spatial resolution as a function of $y_{local}$. The resolution reaches approximately 20~\si{\micro\meter} in the central region and degrades to about 35\dhyphen40~\si{\micro\meter} near the pixel boundaries..}
    \label{fig:uniformity}
\end{figure}

The spatial resolution shows a moderate non-uniformity across the pixel area, as shown in figure~\ref{fig:uniformity}.
The best performance is achieved in the central region of the pixel, where the resolution reaches about 20~\si{\micro\meter}, while it degrades to approximately 35\dhyphen40~\si{\micro\meter} near the pixel boundaries, where the charge-sharing gradients are reduced.

These results demonstrate that the DC-RSD design retains the excellent spatial interpolation capabilities previously achieved with the RSD sensors.
In particular, the parameters extracted from the spatial resolution model are consistent with those reported for optimized RSD geometries ($k_1 = 2.3~\si{\milli\volt}$ and $k_2 = 0.03$)~\cite{MENZIO2024169526}, while benefiting from the improved signal confinement provided by the trench-isolated design.

\section{Timing resolution measurements}
\label{sec:timing_resolution}
\begin{figure}
    \centering
    \includegraphics[width = 0.7\textwidth]{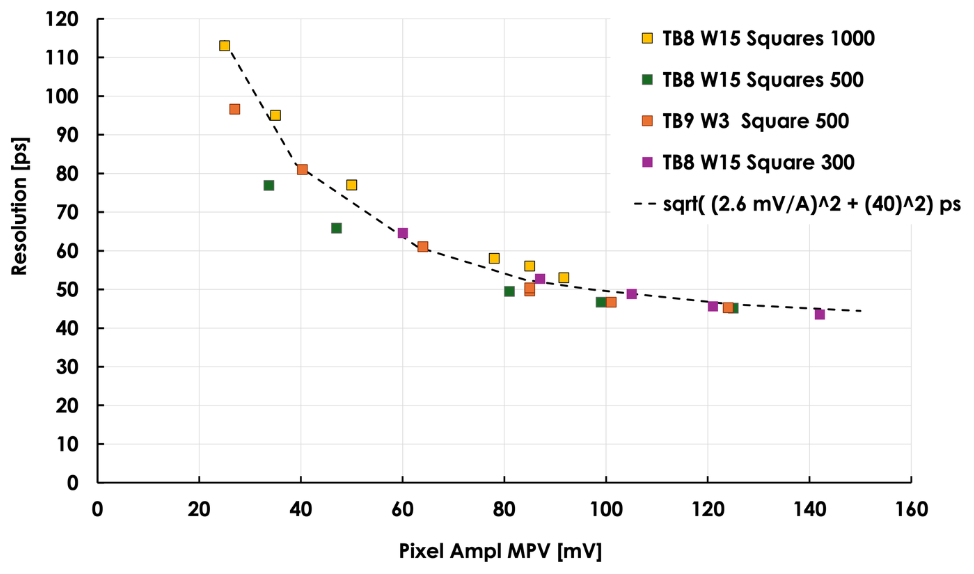}\\
    \caption{Time resolution as a function of the most probable signal amplitude for square DC-RSD matrices with pitches of 300, 500 and 1000~\si{\micro\meter}. The reference detector contribution has been subtracted in quadrature.}
    \label{fig:time_resolution}
\end{figure}

The timing performance of the DC-RSD1 sensors was evaluated using the reconstruction procedure described in Section~\ref{sec:timing}.
The time resolution was measured for square matrices with pitches ranging from 300~\si{\micro\meter} to 1000~\si{\micro\meter} and for different operating gains.
Figure~\ref{fig:time_resolution} shows the measured time resolution as a function of the signal amplitude.
The contribution of the reference detector has been subtracted in quadrature, while the digitizer contribution is not included in the correction.

As expected, the timing resolution improves with increasing signal amplitude owing to the enhanced signal-to-noise ratio.
In contrast to the spatial resolution, no significant dependence on the detector pitch is observed.
Measurements from all investigated geometries follow a common trend, indicating that the timing performance is primarily driven by signal amplitude and electronic noise.

The measured timing resolution can be described as the combination of a jitter-like term and a constant contribution~\cite{ARCIDIACONO2023}:
\begin{equation}
\sigma_t = \sqrt{ \left(\frac{k_1}{MPV}\right)^2 + k_2^2 }
\end{equation}
where $MPV$ is the most-probable pixel amplitude, while the fitted parameters are $k_1 = 2.6~\si{\milli\volt}$ and $k_2 = 40~\si{\pico\second}$, the latter representing the constant term .

For sufficiently large gain values, the sensors achieve time resolutions of approximately 40~\si{\pico\second}, consistent with the $\sim$40~\si{\pico\second} resolutions previously demonstrated with the RSD sensors~\cite{ARCIDIACONO2023}.
These results demonstrate that both AC- and DC-coupled electrodes connected to the resistive layer preserve the excellent timing capabilities characteristic of LGAD-based devices.

\section{Conclusions}
\label{sec:conclusions}
The first FBK production of DC-coupled Resistive Silicon Detectors (DC-RSD1) has been successfully designed, fabricated and characterised through laboratory measurements and beam-test campaigns.
The trench-isolated architecture provides controlled charge sharing, excellent pixel isolation, and fast unipolar signals, addressing some drawbacks of previous, AC-coupled, RSD sensors.
Spatial resolutions of approximately 3\dhyphen4\% of the detector pitch were measured over a wide range of pixel geometries.
Concurrently, time resolutions approaching 40~ps were achieved at high gain, with the timing performance found to depend primarily on the signal-to-noise ratio and not on the detector pitch.
These results validate the DC-RSD concept~\cite{ARCIDIACONO2025170796} as a promising technology for future 4D tracking applications and provide important feedback for the design of the next generation of RSD sensors.

\acknowledgments
The support of the following Funding Agencies is gratefully acknowledged: INFN CSN5 through the 4DSHARE and STARS research project; European Union’s Horizon Europe Research and Innovation program Grant Agreement No. 101057511 (EUROLABS).
The fruitful discussions with the DRD3 collaboration, CERN, are also acknowledged.
The measurements leading to these results have been performed at the test-beam facility at DESY Hamburg (Germany), a member of the Helmholtz Association (HGF), whose support is gratefully acknowledged.


\bibliographystyle{JHEP}
\bibliography{biblio.bib}

\end{document}